\documentclass[fleqn,usenatbib,onecolumn]{mnras}
\usepackage[T1]{fontenc}
\usepackage{ae,aecompl}
\usepackage{graphicx}	
\usepackage{amsmath}	
\usepackage{amssymb}	
\usepackage{comment}
\usepackage{natbib}
\newcommand{\VEC}[1]{\vec {#1} }
\title[Magnetic field in neutron star crust]
{Magnetic field sustained by the elastic force in neutron star crusts}
 \author[Y. Kojima, S. Kisaka, K. Fujisawa]{
  Yasufumi Kojima 
 \thanks{%
 E-mail: ykojima-phys@hiroshima-u.ac.jp}$^1$,
 Shota Kisaka$^1$, 
 Kotaro Fujisawa$^2$\\ 
 $^1${Department of Physics, 
 Graduate School of Advanced Science and Engineering, Hiroshima University,
}\\
{Higashi-Hiroshima, Hiroshima 739-8526, Japan}\\
 $^2${Research Center for Early Universe, Graduate School of Science, 
 University of Tokyo, Bunkyo-ku, Tokyo 113-0033, Japan}
}
\begin{document}
\maketitle 
%
\begin{abstract}
We investigate the magneto--elastic equilibrium of a neutron star crust and magnetic energy stored by the elastic force.
The solenoidal motion driven by the Lorentz force
 can be controlled by the magnetic elastic force,
so that conditions for the magnetic field strength and geometry are less restrictive.
For equilibrium models, the minor solenoidal part
of the magnetic force is balanced by a weak elastic force because
the irrotational part is balanced by the dominant gravity and pressure forces.
Therefore, a strong magnetic field may be confined in the interior, regardless of poloidal or toroidal components.
We numerically calculated axially symmetric models with the maximum shear--strain, and found that a magnetic energy $> 10^{46}$ erg can be stored in the crust, even for a normal surface dipole-field-strength ($<10^{13}$ G).
The magnetic energy much exceeds the elastic energy ($ 10^{44} -10^{45}$ erg).
The shear--stress spatial distribution 
revealed that the elastic structure is likely to break down near the surface.
In particular, the critical position is highly localized
at a depth less than 100 m from the surface. 
\end{abstract}
 
%
\begin{keywords}
  stars: neutron stars: magnetars: magnetic fields
\end{keywords}

\section{Introduction}

Neutron stars are generally observed as radio pulsars, and their magnetic field strength is estimated to be $\sim 10^{12}$ G using the spin period and its derivative.
However, some isolated neutron stars exhibit a field strength higher than $10^{14}$ G. Such neutron stars do not typically
exhibit radio emissions, but emit occasional energetic bursts or anomalous X-ray luminosities.
The rotation rate for such pulsars cannot account for the anomalous emissions\citep[see][for reviews]{2015RPPh...78k6901T,2019RPPh...82j6901E}.
The output energy for such objects can be attributed to strong magnetic fields, and these objects are called magnetars.
However, the surface dipole strength cannot be used to distinguish magnetars and pulsars.
There are some exceptions such as low-field magnetars
$B=6\times 10^{12}- 4\times 10^{13}$ G
\citep{2012ApJ...754...27R,2013ApJ...770...65R,2014ApJ...781L..17R},
and high-magnetic field radio pulsars, 
such as $B=9.4\times 10^{13}$ G in PSR J1847-0130 \citep{2003ApJ...591L.135M}.
Moreover, magnetar-like X-ray outburst from a high-magnetic field radio pulsar PSR J1119-6127 
with $B=4.1\times 10^{13}$ G has been reported previously \citep[][]{2016ApJ...829L..21A}.
The boundary between magnetars and 
high-magnetic field radio pulsars has not been elucidated.
%

The surface dipole field of young neutron stars in some supernova remnants, called central compact objects (CCOs)
\citep{2013ApJ...765...58G,2017JPhCS.932a2006D}, has been estimated
to be $\sim 10^{11}$ G.
Furthermore, their X-ray luminosities exceed the kinetic energy losses, and
are comparable to those of quiescent magnetars \citep{2017ARA&A..55..261K}.
Therefore, a considerable magnetic energy of $\sim 10^{47}$ erg is stored in such objects, 
and the field strength is $\sim 10^{15}$ G in the stellar interior.
Recently, the magneto--thermal evolution of strong and highly tangled magnetic fields in neutron star crusts has been calculated to account for the CCO power
\citep{2020MNRAS.495.1692G,2021ApJ...909..101I,2021ApJ...914..118D}.
%

In some neutron stars, a strong magnetic field $> 10^{14}$ G 
is crucial for various phenomena. However, the locations and 
configurations of such strong fields have not been clarified, and a theoretical understanding of such fields is desired.
Although the stability of magnetized stars has attracted considerable research interest, the underlying phenomenon has not yet been clarified. Simple configurations such as purely toroidal and poloidal magnetic field components are unstable
\citep{1973MNRAS.161..365T,1973MNRAS.163...77M,1973MNRAS.162..339W},
and a twisted-torus configuration is considered a stable equilibrium configuration, in which both components (toroidal and poloidal) stabilize each other.
%

The stability of magnetic stars has been examined numerically
\citep[e.g.,][]{2006A&A...450.1077B,
2011MNRAS.412.1394L,2011MNRAS.412.1730L,2015MNRAS.447.1213M},
following the pioneering work in \cite{2004Natur.431..819B}.
Dynamical simulations revealed 
that the final state after a few Alfv{\'e}n-wave crossing times 
is a twisted-torus configuration, in which 
the poloidal and toroidal components of comparable field strengths tangle with each other.
The fraction of the toroidal component to the total magnetic energy
is greater than $\sim 0.2$
\citep{2009MNRAS.397..763B,2010ApJ...724L..34D,2020MNRAS.495.1360S}.
%

Static or stationary axially symmetric equilibrium models 
have been calculated for various conditions using several methods
\citep[e.g.][]{2005MNRAS.359.1117T,2006ApJS..164..156Y,2006ApJ...651..462Y,
2009MNRAS.395.2162L, 2009MNRAS.397..913C,
2010A&A...517A..58D,2013MNRAS.432.1245F,2013MNRAS.434.2480G,
2015ApJ...802..121A}, and such analyses have been extended to relativistic models
\citep[e.g.][and references therein]{2019PhRvD.100l3019U}.
However, the results reveal that the toroidal magnetic energy in equilibrium is not high.
Instead, special conditions such as
magnetic fields completely confined in deep stellar interiors \citep{2010A&A...517A..58D},
or strong surface currents
\citep{2008MNRAS.385.2080C,
2013MNRAS.432.1245F,2014MNRAS.445.2777F} may be necessary 
to construct models with a large toroidal field.
In addition to the mixed field configuration,
the stratified structure of the density and pressure
is a critical factor for stabilizing such systems
\citep[e.g.,][]{2012MNRAS.420.1263G,2013MNRAS.433.2445A,2019PhRvD..99h4034Y},
as reported in previous studies \citep{1980MNRAS.191..151T}.

Elastic stress, which is another factor in a realistic neutron star model,
has been investigated using dynamic \citep{2020MNRAS.499.2636B}
and static calculations
\citep{2021MNRAS.506.3936K}, and the results show that elasticity is crucial for stability.
In a previous study \citep{2021MNRAS.506.3936K},
we considered the static equilibrium of a magnetized crust
with elastic force, and demonstrated 
that the crust sustained considerable magnetic energy ($> 10^{46}$ erg).
However, we approximated the shear modulus as a
constant throughout the crust.
In this paper, we study the magneto--elastic 
equilibrium in more detail using a realistic shear modulus that changes by a factor of $10^{3}$ in the neutron-star crust.
Moreover, we clarify the spatial dependence of the elastic force and
determine the point where the elastic limit breaks down.
Our results will be of astrophysical relevance in studying abrupt bursts and magnetic field evolution in a secular timescale. 
%

The remainder of this manuscript is organized as follows.
Section 2 presents the models and equations relevant to our problem.
The shear modulus is assumed to be proportional to density, and the elastic force is formulated. The numerical results are presented in Section 3. Finally, Section 4 concludes this paper.
%

\section{Mathematical formulation}
\subsection{Force balance}

We consider an equilibrium state for the crust of a magnetized neutron star and use spherical coordinates $(r, \theta, \varphi)$ to represent an axially symmetric configuration 
$\partial_{\varphi}=0$.
The static force balance for a nonrotating star
in the Newtonian gravity is given by
\begin{equation}
-{\VEC \nabla}P-\rho{\VEC \nabla}\Phi_{\rm G}
+ {\VEC f}+{\VEC h} =0,
  \label{Forcebalance.eqn}
\end{equation}
where the first two terms expressed by pressure $P$,
mass density $\rho$, and gravitational potential $\Phi_{\rm G}$ are dominant.
We further assume a barotropic distribution, $P=P(\rho)$,
and the sum of these forces is expressed by $ -\rho {\VEC \nabla}\Phi_{\rm eff}$. The resultant stellar structure is spherically symmetric.
The barotropic approximation restricts the configuration such that no stable equilibrium exists
\citep[e.g.][]{2009A&A...499..557R,2012MNRAS.424..482L}.
However, we adopt the barotropic approximation to examine
the effect of the elastic force in a simplified setup.
Otherwise, a solenoidal acceleration
${\VEC{\nabla}}\times (\rho^{-1} {\nabla}P)\ne 0$ is generated
in a non-barotropic star.
%

The third term in eq.(\ref{Forcebalance.eqn})
is the Lorentz force ${\VEC f} \equiv c^{-1}{\VEC j}\times {\VEC  B}$, 
which is considerably smaller than the dominant forces
and different in nature.
Typically, the Lorentz force causes non-radial acceleration.
The static force-balance with the Lorentz force constrains the field strength and configuration of the magnetic field. For example, 
the energy ratio of the toroidal to poloidal components
is limited to small values for barotropic magnetohydrodynamic (MHD) equilibrium, as discussed in Section 1.
%

This constraint can be relaxed considerably by incorporating the elastic force ${\VEC h}$, irrespective of its
small magnitude \citep{2021MNRAS.506.3936K}.
The Lorentz force comprises irrotational and solenoidal parts.
The irrotational force is easily balanced by the dominant forces, but 
the solenoidal force is never balanced.
However, the elastic force can be used to balance 
the solenoidal component of the Lorentz force.
%

Equation(\ref{Forcebalance.eqn}) cannot be directly solved for
magneto--elastic equilibrium models, because the order of their strengths differs considerably;
the Lorentz ${\VEC f}$ and elastic ${\VEC h}$ forces
are considerably smaller than the gravity and pressure forces.
An approximation is given by
\begin{equation}
{\VEC f}+{\VEC h}=0.
\label{strongbal} 
\end{equation}
The weak forces are balanced under fixed stellar structures.
Instead of eq.(\ref{strongbal}), we use a weaker constraint.
Thus, in the solenoidal part, the "curl" of eq.(\ref{Forcebalance.eqn})
should be balanced between ${\VEC f}$ and ${\VEC h}$, because
the irrotational part may be balanced by a small perturbation in
the pressure and gravity. A set of approximated equations is as follows:
\begin{equation}
({\VEC f}+{\VEC h})_{\varphi}=0, 
   \label{blanceeqn3a}
\end{equation}
\begin{equation}
   [{\VEC{\nabla}}\times \rho^{-1}({\VEC f}+{\VEC h})]_{\varphi}=0.
\label{rotforce3} 
\end{equation}
We consider the azimuthal component only in eq.(\ref{rotforce3}), because
the poloidal components vanish, as evident from eq.(\ref{blanceeqn3a}) and the axial symmetry ($\partial_{\varphi}=0$).
%

\subsection{Elastic force in the crust}

Our consideration of the magnetic--elastic equilibrium in the solid crust of a neutron star is limited to the inner crust,
where the mass density $\rho$ ranges from 
$\rho_{c} =1.4\times 10^{14}$ g cm$^{-3}$ 
at the core--crust boundary $r_{c}$ 
to the neutron-drip density $\rho_{1} = 4\times 10^{11}$ g cm$^{-3}$
at $R$.
We ignore the outer crust, which
could be considerably affected in the presence of a strong magnetic field, and assume the exterior region of $r> R$ as the vacuum.
The spatial profile for $r_c \le r \le R$ is approximated as
\citep{2019MNRAS.486.4130L,2020MNRAS.494.3790K}.
\begin{equation}
\frac{\rho}{\rho_{c}}
=\left[1-\left(1-\left(\frac{\rho_1}{\rho_{c}}\right)^{1/2}
 \right)\left(\frac{r-r_{c}}{d} \right)\right]^2 ,
 \end{equation}
where $d=R-r_c$ is the crust thickness, assumed to be $d/R=0.1$.
%

The shear modulus $\mu$ increases with the density, and may be
approximated as a linear function of $\rho$,
which is overall fitted to the results of a detailed calculation reported previously
\citep[see Fig.~43 in ][]{2008LRR....11...10C}.
$\mu$ is expressed as
 \begin{equation}
    {\mu}=\frac{\mu_{1}\rho}{\rho_{1}} ,
 \end{equation}
where $\mu_{1}=3 \times 10^{27}{\rm erg~cm}^{-3}$ at the
outer boundary. The value increases with $\rho$ until 
$ \mu_{c} \approx 10^{30}{\rm erg~cm}^{-3}$ at
the core--crust interface.

The elastic force ${\VEC h}$ in the solid crust is considered, 
and the {\it i}~th component $h_{i}$ is expressed
using the trace-free strain tensor $\sigma_{ij}$ and $\mu$ as
\begin{equation}
h_{i}= {\nabla}_{j} \left(\mu \sigma^{j}_{i} \right),
 \label{ElasticForce.eqn}
\end{equation}
where $\sigma_{ij}$ is expressed in terms of the displacement vector $\xi_{i}$ and three-dimensional metric $g_{ij}$ as
\begin{equation}
 \sigma_{ij} ={\nabla}_{i} \xi _{j}  +{\nabla}_{j} \xi _{i}
-\frac{2}{3} g_{ij} ({\nabla}_{k} \xi ^{k}).
\end{equation}
The components of eq.(\ref{ElasticForce.eqn})
in spherical coordinates is expressed as
\begin{eqnarray}
    h_{r}&=& \left[
 -{\VEC{\nabla}}\times \left(\mu
   {\VEC{\omega}}_{\varphi} \right)
 + 2 {\VEC{\nabla}}( \mu {\VEC{\nabla}}\cdot{\VEC{\xi}} )
 \right]_{r}
 -2\mu_{,r}
 \left[ \frac{2}{r}\xi_{r}
 +\frac{1}{\varpi}
 (\sin\theta\xi_{\theta})_{,\theta}
 \right]
  +\frac{2\mu_{,\theta}}{r}  \xi_{\theta,r},
\label{elshr.eqn}
  \\
    h_{\theta}&=& \left[
 -{\VEC{\nabla}}\times \left(\mu
  {\VEC{\omega}}_{\varphi}\right)
+ 2 {\VEC{\nabla}}( \mu {\VEC{\nabla}}\cdot{\VEC{\xi}} )
 \right]_{\theta}
 +\frac{2\mu_{,r}}{r}
 (\xi_{r,\theta}-\xi_{\theta})
 -\frac{2\mu_{,\theta}}{r\varpi}
 \left[\cos\theta \xi_{\theta}
 +\sin \theta(r\xi_{r} )_{,r} 
 \right],
\label{elsht.eqn}
 \\
    h_{\varphi}&=& \frac{1}{\varpi^{3}}\left[\mu \varpi^{4}
  \left(\frac{\xi_{\varphi}}{\varpi}\right)_{,r}\right]_{,r}
 +\frac{1}{r^{2}\varpi^{2}}\left[\mu \varpi^{3}
 \left( \frac{\xi_{\varphi}}{\varpi}\right)_{,\theta}\right]_{,\theta} ,
 \label{xi3.eqn}
\end{eqnarray}
where 
\begin{equation}
  {\VEC{\omega}}_{\varphi}  
={\VEC{\nabla}}\times {\VEC \xi}_{p}
=\frac{W}{\varpi} \VEC{e}_{\varphi},
 \label{omg3byxip.eqn}
\end{equation}
and $\varpi=r\sin\theta$ is the cylindrical radius; 
$\VEC{e}_{\varphi}$ is the unit vector along the azimuthal direction. 
In eq.(\ref{omg3byxip.eqn}), the azimuthal vector ${\VEC{\omega}}_{\varphi}$ is expressed as a scalar function $W$.

 We limit our consideration to incompressible motion and introduce a function $X$ to express ${\VEC{\xi}}_{p}$ as given below,
which satisfies the condition ${\VEC{\nabla}}\cdot{\VEC{\xi}}=0$,
\begin{equation}
{\VEC \xi}_{p} = {\VEC{\nabla}}\times \left(\frac{X}{\varpi} {\VEC e}_{\varphi}\right).
\end{equation}
Equation (\ref{omg3byxip.eqn}) is reduced to 
a relation between $X$ and $W$ as
\begin{equation}
\left[ {\VEC{\nabla}}\times {\VEC{\nabla}}\times 
\left( \frac{X}{\varpi}{\VEC e}_{\varphi}\right) \right]_{\varphi} 
= \frac{W}{\varpi}.
\label{XWrelation.eqn}
\end{equation}
Under the approximation $\mu=\mu(r)$, 
the elastic acceleration term in eq.(\ref{rotforce3})
with eqs.(\ref{elshr.eqn}) and (\ref{elsht.eqn}) is expressed by
\begin{eqnarray}
({\VEC{\nabla}}\times \rho^{-1}{\VEC h})_{\varphi} 
  = -\left[ {\VEC{\nabla}}\times \left( \rho^{-1}
 {\VEC{\nabla}}\times
 \left( \frac{\mu W}{\varpi} {\VEC e}_{\varphi}\right) \right)
 \right]_{\varphi} 
 +\frac{2}{\varpi}
  \left[\left( \frac{\rho^{-1}\mu^{\prime}}{r}X_{,r} \right)_{,r}
  +\frac{(\rho^{-1}\mu^{\prime})^{\prime}\sin\theta}{r^2}
\left( \frac{1}{\sin\theta}
  X_{,\theta}\right)_{,\theta}
 \right],
 \label{rotWWaa.eqn}
\end{eqnarray}
where a prime $^\prime$ denotes derivative with respect to $r$.
%

\subsection{Magnetic field}

An axially symmetric magnetic field is described using two functions as
\begin{equation}
\VEC{B}=
\VEC{\nabla}\times \left(
\frac{\Psi}{\varpi}\VEC{e}_{\varphi}\right)
+\frac{S}{\varpi}\VEC{e}_{\varphi} ,
\label{eqnDefBB}
\end{equation}
and the electric currents are derived from $\VEC{B}$ using the Amp{\'e}re--Biot--Savart law
\begin{equation}
\frac{4\pi}{c}\VEC{j} =
\VEC{\nabla}\times \left(
\frac{S}{\varpi}\VEC{e}_{\varphi}\right)
+
\VEC{\nabla}\times \VEC{\nabla}\times \left(
\frac{\Psi}{\varpi}\VEC{e}_{\varphi}\right).
\label{ampbiosav.eqn}
\end{equation}
The current function $S$ is a function of $\Psi$,  when the azimuthal component of the Lorentz force vanishes, $f_{\varphi}=0$.
An irrotational condition
$\VEC{\nabla}\times (\rho^{-1}  {\VEC f})=0$
constrains the azimuthal current as
\begin{equation}
\frac{4\pi\varpi}{c}j_{\varphi} = -\rho K^\prime \varpi^2 +S^\prime S,
 \label{MHDeqil.eqn}
\end{equation}
where $K$ is a function of $\Psi$, and
a prime $^\prime$ denotes the derivative with respect to $\Psi$.
Thus, the Lorentz force is reduced to
\begin{equation}
 {\VEC f}=- \frac{\rho K^\prime}{4\pi}{\VEC \nabla}\Psi.
\end{equation}
%

A simple model of barotropic MHD equilibrium
is obtained by assuming that the field is purely 
dipolar ($\Psi \propto \sin^2\theta$ and $S=0$),
and that $K^\prime$ is a constant.
We assume that the magnetic field exists outside the core, and
the field inside the crust is smoothly connected to the external dipole in vacuum.
For the magnetic field in barotropic MHD equilibrium, a solenoidal force does not exist, i.e., the
elastic force vanishes, ${\VEC \xi}=0$.
We adopt this magnetic field configuration as a reference model and examine the response of the elastic force (${\VEC \xi}\ne0$) by adding another arbitrary field. We consider the possibility of a strong field confined inside the crust,
irrespective of whether the component is poloidal or toroidal. 
The choice of the models is described as follows, and is listed in Table~\ref{table1:mylabel}.

\begin{table}
    \centering
    \begin{tabular}{cll}
Name &  Magnetic field geometry & Elastic motion\\
\hline 
 P & 
      Confined poloidal field expressed by eq.(\ref{loopPsi}) & ${\VEC \xi}_{p}$
   \\
  T &
     Confined toroidal field expressed by eq.(\ref{loopB3}) & ${\VEC \xi}_{p},  \xi_{\varphi}$
   \\
 M &
       Mixed poloidal--toroidal field using eqs.(\ref{loopPsi}) and (\ref{loopB3}),
      & ${\VEC \xi}_{p},  \xi_{\varphi}$
    \\ \hline
    \end{tabular}
    \caption{Description of the magnetic field for the equilibrium model with elastic force}
\label{table1:mylabel}
\end{table}

The poloidal magnetic field confined in the crust ($r_{c}\le r\le R$)
may be modeled as
\begin{equation}
\Psi_{(1)}=N_{1}B_{0}R^{-2}[(r-r_c)(r-R)]^2\sin{\theta}P_{l}^{\prime}, 
\label{loopPsi}
\end{equation}
where $P_{l} (\cos \theta)$ is the Legendre polynomial with multipole index $l$, and a prime $^\prime$ denotes the derivative with respect to 
$\cos \theta$.
Similarly, the confined toroidal field $B_{\varphi}=S/\varpi$ 
may be modeled as
\begin{equation}
S_{(2)}=N_{2}B_{0}R^{-3}[(r-r_c)(r-R)]^2 \sin{\theta}P_{l}^{\prime}. 
\label{loopB3}
\end{equation}
In eqs.(\ref{loopPsi}) and (\ref{loopB3}), 
we use $B_{0}\equiv|B_{0}(R,0)|$
for the reference field ${\VEC{B}}_{0}$ at the pole
as the normalization of the magnetic field strength; 
$N_{s} (s=1,2)$ is a dimensionless number.
Here, the radial function is selected such that it vanishes at the core--crust interface and at the surface.
The electric current ${\VEC j}_{({\rm s})}$ 
$(s=1,2)$ is calculated 
from $\Psi_{(1)}$ or $S_{(2)}$ using eq.(\ref{ampbiosav.eqn}).

Three equilibrium sequences are considered, namely models P, T, and M. Model P represents a purely poloidal magnetic field, for which the elastic motion ${\VEC {\xi}}_{p}$ is solely driven by the Lorentz force.
 The magnitude $|{\xi}_{p}|$ increases with $|N_{1}|$.
Model T represents a purely toroidal magnetic field, for which both components ${\VEC {\xi}}_{p}$ and $\xi_{\varphi}$ are driven by the Lorentz force.
The magnitude $|\xi|$ also increases with $|N_{2}|$.
Model M represents a mixed model of poloidal and toroidal magnetic fields.
The magnitude $|\xi|$ depends on $N_{1}$ and $ N_{2}$, and the ratio $|N_{2}/N_{1}|$ determines the dominant component.
We fix the ratio as $|N_{2}/N_{1}|=10^2$, for which the typical magnetic field strength is given by $|B_{\varphi}|/|B_{{\rm p}}| \sim 10$.
%

A critical limit of normalization is determined by
either $N_{1}$ or $ N_{2}$, beyond which the crust does not respond elastically.
We discuss the explicit condition in the next subsection.
In Section 3, we show the numeric calculation of a state with the maximum strain, characterized by a quantity $\sigma_{c}$.
These models depend on the multipole index $l$
for the additional magnetic field and are summarized in Table~\ref{table1:mylabel}.

\subsection{Elastic limit and parameter}
No elastic shear-deformation ${\VEC \xi}=0$
exists in the reference configuration, where
the magnetic field and current are denoted as
${\VEC B}_{(0)}$ and ${\VEC j}_{(0)}$, respectively. 
We calculate ${\VEC \xi} $, which responds to the Lorentz force
${\VEC{f}}=c^{-1}{\VEC{j}} \times {\VEC{B}}$
for ${\VEC{j}} = {\VEC{j}}_{(0)}+{\VEC{j}}_{(s)}$
and
${\VEC{B}} = {\VEC{B}}_{(0)}+{\VEC{B}}_{(s)}$,
($s=1,2$).
The displacement $|\xi|$ in the equilibrium model increases with the normalization constant $|N_{{\rm s}}|$ in 
eqs.(\ref{loopPsi}) and (\ref{loopB3}).
We adopt the von Mises criterion\citep[e.g.,][]{
    1969imcm.book.....M,2000MNRAS.319..902U}
  to estimate the maximum value of $|\xi|$, or the shear tensor $|\sigma_{ij}|$.
The numerical calculation adopting the von Mises criterion is described in the next subsection.
The criterion is empirical formula based on the energy. Another criteria, the Tresca, 
is based on the difference between the maximum and minimum of shear.
   Two criteria do not predict the same critical states
  \citep[e.g.,][]{1969imcm.book.....M}. 
Our concern is the order-of-magnitude level.  
The Mises criterion is expressed as follows:
\begin{equation}
\frac{1}{2}\sigma_{ij}\sigma^{ij}
\le (\sigma_{c})^{2},
   \label{criterion}
\end{equation}
where $\sigma_{c}$ is the maximum strain with a definite value,
$\sigma_{c} \approx 10^{-2}-10^{-1}$
\citep{2009PhRvL.102s1102H,2018PhRvL.121m2701C,2018MNRAS.480.5511B}. 
%

The elastic limit (\ref{criterion}) thus constrains the normalization 
constant $N_{{\rm s}}$.
We focus on the equilibrium model with the maximum strain $\sigma_{c}$.
The calculated model depends on the ratio $q$ of the elastic and Lorentz forces.
\begin{equation}
 q\equiv\frac{4 \pi \mu_{1} \sigma_{c}}{B_{0}^2}
\approx 0.4
  \left(\frac{B_{0} }{10^{14}{\rm G}}  \right)^{-2}
  \left(\frac{ \sigma_{c}}{0.1}\right).
\label{parameter.eqn}
\end{equation}
Here, $B_{0}$ is the magnetic field strength at the pole;
the shear modulus, $\mu_{1}= 3\times 10^{27} {\rm erg~cm}^{-3}$, at the surface is used.
The limit $q=0$ corresponds to our reference model. 

The elastic energy stored inside a crust of volume $V$ is given by
\begin{equation}
E_{\rm ela}=\int_{V} \frac{\mu}{2} \sigma_{ij} \sigma^{ij}  dV.
    \label{Elsenergy.eqn}
\end{equation}
For numerical integration, we fix the stellar model with 
radius $R = 12$ km and crust thickness $d = 0.1R$, and the lowest shear modulus
$\mu_{1}= 3\times 10^{27}{\rm erg~cm}^{-3}$.
The magnetic energy $E_{\rm mag}$ is given by
\begin{equation} 
E_{\rm mag} =\int_{V} \frac{1}{8\pi} B^2 dV.
 \label{Magenergy.eqn} 
\end{equation}
The elastic energy is scaled to 
$E_{\rm ela} \propto  \sigma_{c}^2$
for obtaining an equilibrium model with the maximum strain,
whereas the magnetic energy is scaled to $ E_{\rm mag} \propto  B_{0}^2$.
The ratio 
$ E_{\rm ela}/ E_{\rm mag} \propto \sigma_{c}^2 B_{0}^{-2}$
with respect to the energy is not equivalent to
that of their forces, i.e., $q \propto \sigma_{c} B_{0}^{-2}$. 
%

\subsection{Numerical method}
We numerically calculate the elastic displacement ${\VEC{\xi}}$ against the Lorentz force in the crustal region $r_{c} \le r \le R$ with thickness $d/R=(R-r_{c})/R=0.1$.
In general, the Lorentz force for 
${\VEC B}_{(0)}+{\VEC B}_{({\rm s})} $ and 
${\VEC j}_{(0)} +{\VEC j}_{({\rm s})}$ 
results in nonzero source terms in 
eqs.(\ref{blanceeqn3a}) and (\ref{rotforce3}), respectively. 
We examine the elastic response 
by solving a set of partial differential equations
(\ref{blanceeqn3a}) and (\ref{rotforce3}) 
with eqs.(\ref{xi3.eqn}), (\ref{XWrelation.eqn})
and (\ref{rotWWaa.eqn}).
To solve these equations, we use expansion via Legendre polynomials
$P_l(\cos(\theta))$ and radial functions
$x_l(r)$, $w_l(r)$, and  $k_l(r)$ as
\begin{eqnarray}
  X&=& -\sum_{l\ge 1} x_{l}\sin \theta P_{l,\theta},
    \\
    \mu W&=& -\sum_{l\ge 1} w_{l}\sin \theta P_{l,\theta},
    \\
\xi_{\varphi}&=& -\sum_{l\ge 1} r k_{l} P_{l,\theta}.  
\label{xi3lp.eq28}
\end{eqnarray}
The source terms derived from the Lorentz force
are expanded with radial functions $a_{l}(r)$ and $b_{l}(r)$ as
\begin{equation}
  f_{\varphi}= -\sum_{l\ge 1}  r^{-3} a_{l} P_{l,\theta}.  
 \label{expndf3.eqn}
\end{equation}
\begin{equation}
[{\VEC{\nabla}}\times \rho^{-1}{\VEC{f}} ]_{\varphi}
= -\sum_{l\ge 1}  r^{-1} b_{l} P_{l,\theta}.
\label{expndLF.eqn}
\end{equation}
%

By using eqs.(\ref{xi3.eqn}), (\ref{xi3lp.eq28}), and
(\ref{expndf3.eqn}), eq.(\ref{blanceeqn3a}) for 
the azimuthal component, $\xi_{\varphi}$,
is deduced to an ordinary differential equation
for $k_{l} ~(l=1,2,\cdots)$;
\begin{equation}
 (\mu r^{4} k_{l}^{\prime})^{\prime}
    -(l-1)(l+2)\mu r^{2} k_{l}
    =-a_{l},
    \label{klexpd.eqn} 
\end{equation}
where the orthogonality of the Legendre polynomials is used.
Similarly, eq. (\ref{XWrelation.eqn}) for the 
functions $X$ and $W$ becomes 
\begin{equation}
    x_{l}^{\prime\prime}
    -\frac{l(l+1)}{r^2}x_{l}+\frac{1}{\mu}w_{l}=0 .
    \label{flexpd.eqn}
\end{equation}
Equation (\ref{rotforce3}) with eqs.(\ref{rotWWaa.eqn}) and (\ref{expndLF.eqn}) is also deduced to 
\begin{equation}
(\rho^{-1}w_{l}^{\prime})^{\prime}
    -\left( \frac{2\mu^{\prime}}{\mu r} +\frac{l(l+1)}{r^2}
    \right) \rho^{-1}w_{l}
   +2\left(\frac{\mu^{\prime}}{\rho r}
    \right)^{\prime}
   \left(x_{l}^{\prime} -\frac{l(l+1)}{r}x_{l}
    \right)
    =-b_{l}.
    \label{glexpd.eqn}
\end{equation}
In deriving eq.(\ref{glexpd.eqn}), $x_{l}^{\prime\prime}$ is originally used in eq.(\ref{rotWWaa.eqn}), 
but is eliminated by eq.(\ref{flexpd.eqn}).
The displacement ${\VEC{\xi}}$ is decoupled 
with respect to the index $l$
because of the spherical symmetry, i.e.,
$\rho(r)$ and $\mu(r)$. 
The analytic expressions for $a_{l}(r)$ and $b_{l}(r)$ in the source terms are complicated.
We numerically evaluate the
coefficients $a_{l}(r)$ and $b_{l}(r)$ in eqs.(\ref{expndf3.eqn}) and (\ref{expndLF.eqn}) for the Lorentz force originated from 
${\VEC B}_{(0)}+{\VEC B}_{({\rm s})} $ and 
${\VEC j}_{(0)} +{\VEC j}_{({\rm s})}$ by using the orthogonality of the Legendre polynomials.
In the numerical calculation, we truncated the number of
$l$, and determined that the maximum value of $l$ in the range of 40--60 is sufficient for numerical convergence, because the source terms are specified by low multipole $l$ in eqs.(\ref{loopPsi}) and (\ref{loopB3}).
%

We discuss the boundary conditions for these radial functions $k_{l}$, 
$x_{l}$, and $w_{l}$.
The boundary conditions for the axial component $\xi_{\varphi}$
are $\xi_{\varphi}=0$ at $r=r_c$ and $\sigma_{r\varphi}=0$, which is $r(\xi_{\varphi}/r)_{,r}=0$ at $r=R$.
Therefore, we impose $k_{l}=0$ at $r_c$ and $k_{l}^{\prime}=0$ at $R$.
The direction of the poloidal displacement is horizontal at the radial boundaries.
Thus, $x_{l}=0$ at both $r_c$ and $R$.
Finally, because the shear modulus
does not exist outside the crust,
$\mu W $ approaches zero at the radial boundaries. Therefore, we impose
$w_{l}=0$ at both $r_c$ and $R$.

By increasing the normalization constant $|N_{{\rm s}}|$ in
eqs.(\ref{loopPsi}) and (\ref{loopB3}),
the source terms $a_{l}$ and $b_{l}$ 
in eqs.(\ref{klexpd.eqn}) and
(\ref{glexpd.eqn}) also increase in magnitude.
The maximum displacement $|\xi|$ or the maximum shear-tensor $|\sigma_{ij}|$ determined by the criterion (\ref{criterion}) limits $|N_{{\rm s}}|$.
 The critical value $|N_{{\rm s}}|$
 is determined by numerical solutions of
  eqs.(\ref{klexpd.eqn}), (\ref{flexpd.eqn}), and (\ref{glexpd.eqn}).
  The value is relevant to the maximum energy in the elastic limit, whereas
  it does not affect the spatial distribution of the shear $\sigma_{ij}$, 
  which depends on the magnetic field configuration.

\section{Results}
\subsection{Energy}

\begin{figure}
\centering
\includegraphics[scale=0.95]{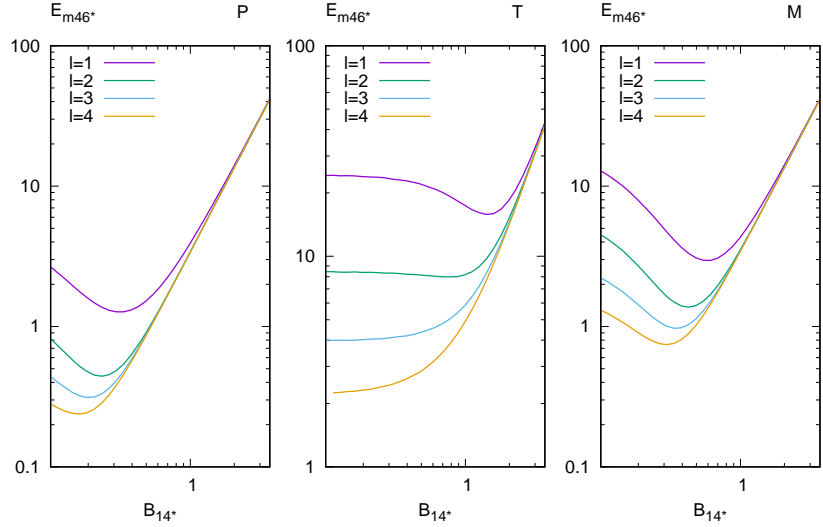}
%
\caption{Magnetic energy as a function of the surface dipole field strength
for the three models listed in Table \ref{table1:mylabel}
for various multipole $l$ values.
}
\label{fig:EBvsBS}
\end{figure}
  
The magneto--elastic equilibrium with maximum strain is 
 specified using the dimensionless 
 parameter $q$ in eq.(\ref{parameter.eqn}), which is
 the ratio of the elastic and magnetic forces.
We calculate various equilibrium models in the range of $q$ $~(10^{-2.5} \le q \le 10^{1.5})$.
We first discuss how the magnetic energy (\ref{Magenergy.eqn})
depends on the surface dipole field strength.
The energy $E_{\rm mag}$ depends on $B_{0}^2$, but we eliminate 
$B_{0}^2$ by $q$, so that 
$E_{\rm mag}$ is expressed by $E_{{\rm m}46*}$ in 
the ordinate of Fig.~\ref{fig:EBvsBS}, where
\begin{equation}
 E_{\rm mag}
   \equiv  E_{{\rm m}46*} (\sigma_{c} /0.1)
   \times 10^{46}~{\rm erg}.
\end{equation}
Similarly, the abscissa of Fig.~\ref{fig:EBvsBS}
is $B_{14*}$, which is related to $B_{0}$ as
\begin{equation}
B_{0} \equiv  B_{14*} (\sigma_{c} /0.1)^{1/2}
 \times 10^{14}~{\rm G}.
\end{equation}
Parameter $q$ in eq.(\ref{parameter.eqn}) is given by 
$q \approx 0.4  (B_{14*})^{-2}$.
Figure~\ref{fig:EBvsBS} displays our numerical results for the three models listed in Table~\ref{table1:mylabel}
for various multipole $l$ values.
These results indicate that there exist two branches of equilibrium models.
This property is common for all magnetic geometries,
regardless of whether the component is poloidol or toroidal.
One model features a high field strength of the surface dipole,
which corresponds to $B_{14*}> 1$ (equivalent to $q\ll 1$).
The magnetic energy is proportional to $B_{14*}^2$.
For the models in this branch, the elastic force is too weak to
sustain the magnetic field deviating from ${\VEC B}_{(0)}$, and thus
the magnetic field is the same as that of a barotropic MHD model.

Equilibrium models in the other branch, corresponding to 
$B_{14*}< 1$ ($q\gg 1$), exhibit different properties.
The magnetic energy inside the crust increases 
in Models P and M and approaches a constant in Model T
with decrease in $B_{14*}$.
A significantly high magnetic energy
$\gtrsim 10^{46}$ erg is supported by the elastic force,
even for a surface dipole field strength of $\sim 10^{13}$ G.
This result is explained as follows.
The maximum strain is located near the surface in Model P.
The additional field ${\VEC B}_{(s)}$ always exists inside the crust
and approaches zero toward the surface.
By reducing the strength of the penetrating field ${\VEC B}_{(0)}$, 
the critical condition near the surface is relaxed, so that
the amount of magnetic energy associated with ${\VEC B}_{(s)}$
stored deep inside the crust increases.
The maximum strain is located at the middle of the crust in Model T.
The dependence of $B_{14*}$ on the magnetic energy is less
sensitive.

The boundary of the two classes is approximately
$B_{14*}=0.3$, which corresponds to a physical value
 $B_{0}=3\times 10^{13} (\sigma_{c} /0.1)^{1/2}{\rm G}$.
The critical value is presented by $q \sim 4$ in eq.(\ref{parameter.eqn}).
Two distinct classes are demonstrated in the constant $\mu$ models
\citep{2021MNRAS.506.3936K},
and the results are consistent with those of previous studies
for suitable choices of $q$.

Extrapolating our models to either small or large values of $B_{14*}$ 
results in further increase in the magnetic energy $E_{\rm mag}$ in the crust. 
The energy is several times $10^{47}$ erg, and
the field strength correspondingly exceeds $10^{15}$ G.
In such an equilibrium model,
the magnetic force becomes comparable to the dominant forces near the surface,
and our perturbation approach becomes inadequate.
These results are applicable to a range of
$E_{\rm mag} \le E_{c}$, where $E_{c}$ is a few times $10^{47}$ erg.
%

\begin{figure}
\centering
\includegraphics[scale=0.95]{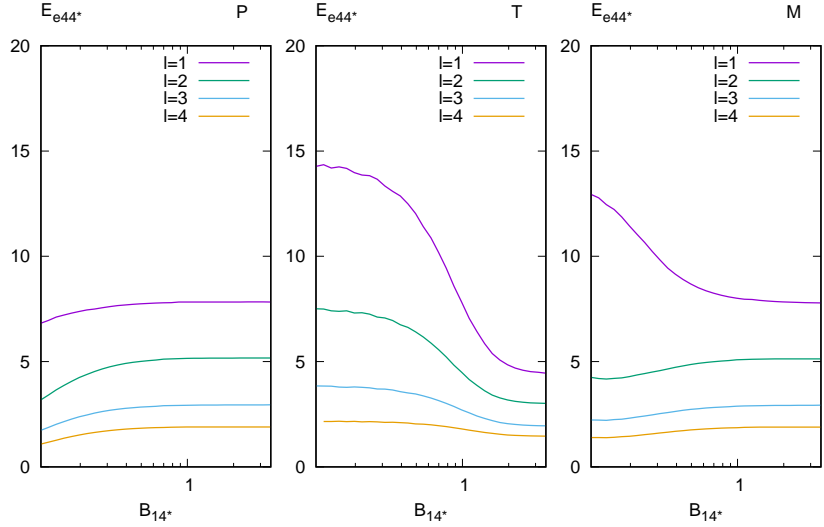}
%
\caption{Elastic energy as a function of the surface 
dipole field strength for the three models listed
in Table \ref{table1:mylabel} for various multipole $l$ values.
}
\label{fig:ELvsBS}
\end{figure}

Figure~\ref{fig:ELvsBS} shows
the elastic energy $E_{\rm els}$ by $E_{{\rm e}44*}$, where
\begin{equation}
    E_{\rm els}
\equiv E_{{\rm e}44*}  (\sigma_{c} /0.1)^{2} \times 10^{44}~{\rm erg}.
\end{equation}
 The elastic energy changes by a factor of 3,
even for varying $q$ in the range of $10^{-2.5} \le q \le 10^{1.5}$.
The elastic energy is approximately constant, because
all equilibrium models correspond to the maximum shear--strain,
where the strength is fixed by the von Mises criterion (\ref{criterion}). However, the spatial distribution of $\sigma_{ij}$ changes slightly
owing to the combined magnetic fields.
The dependence on $l$, i.e., a decrease in energy with $l$
in Fig.~\ref{fig:ELvsBS}, 
mainly originates from the volume integration with the node number $l$,
whereas the maximum is fixed irrespective of $l$. 
%

 The elastic energy is $E_{\rm els} \sim 10^{44} -10^{45}$ erg
for all models with $\sigma_{c} =0.1$. 
The magnetic energy $E_{\rm mag} \sim 10^{45} -10^{47}$ erg in 
Fig.~\ref{fig:EBvsBS} always exceeds the elastic energy $E_{\rm els}$.
As discussed previously
\citep{2021MNRAS.506.3936K},
this fact can be attributed to a large amount of magnetic energy 
associated with the irrotational part of the magnetic field, 
which is balanced by the gravity and pressure. 
Thus, for the equilibrium models, the minor solenoidal 
component may be balanced by a weak elastic force.
Therefore, a magnetic energy larger than the elastic energy
is allowed in these models.

\subsection{Magnetic field and shear stress}

\begin{figure}
\centering
    \includegraphics[scale=1.2]{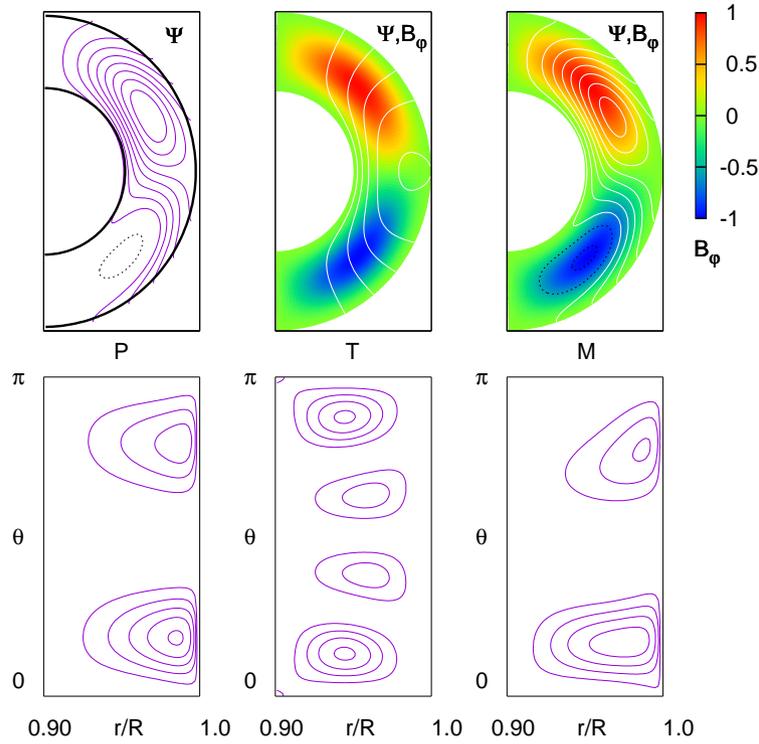}
\caption{
\label{FigBSHx3}
Magnetic field (top panels) and magnitude of shear stress
$\sigma^{ij}\sigma_{ij}/(2\sigma_{c}^2)$ (bottom panels)
for Models P, T, and M listed in Table \ref{table1:mylabel},
from left to right.
The contour of $\Psi$ in the top panels is plotted 
With an interval of $\Delta \Psi/(B_{0}R^2) = 0.1$
in spherical coordinates "enlarged" in the radial direction for display purposes.
The dotted lines represent the negative value of the contour level. The
color contour represents the toroidal field $B_{\varphi}$
normalized by the maximum value.
The contour of $\sigma^{ij}\sigma_{ij}/(2\sigma_{c}^2)$ in the $r$-$\theta$ plane
is plotted with $\Delta =0.15$ 
for the crustal region.
}
\end{figure}

Figure~\ref{FigBSHx3} displays the magnetic field for the
three models listed 
in Table \ref{table1:mylabel}
with $l=2$ and $q=10$, which corresponds to $B_{14*} \approx 0.2$.
The crustal thickness in the figure is enlarged by five times for suitable display.
The maximum of the field strength is 
$B_{p}/B_{0}\sim 31$ in Model P (left panel);
$B_{p}/B_{0}\sim 22$,
$B_{\varphi}/B_{0}\sim 114$ in Model T (middle panel);
and $B_{p}/B_{0}\sim 28$,
$B_{\varphi}/B_{0}\sim 61$ in Model M (right panel).
These fields change considerably from the
purely dipolar field ${\VEC{B}}_{0}$, whose magnetic function 
is the same as that of the middle panel.
The poloidal field is confined in Model P, and the
strong toroidal field is confined in Models T and M. 
Model M may be regarded as a combination of Models P and T.
%

Elastic displacements are induced in these models,
and the magnitude of shear $\sigma^{ij}\sigma_{ij}/(2\sigma_{c}^2)$ 
is displayed in the bottom panels of Fig.~\ref{FigBSHx3}.
The poloidal component ${\vec{\xi}_{p}}$ is only induced in Model P, and
the dominant shear component is found to be  
$\sigma_{r\theta}=r(r^{-1} \xi_{\theta})_{,r}+r^{-1} (\xi_{r})_{,\theta}$.
Two peaks originate from the additional magnetic field 
in which the angular function is specified by $l=2$, and the radial function has no node.
The position of the maximum shear--strain 
is not at the center of the crust but is shifted near the surface $r/R \sim 0.99$.

In models with the toroidal component, the axial component 
$\xi_{\varphi}$ is also induced.
The shear component 
$\sigma_{r \varphi}=r(\xi_{\varphi}/r)_{,r} \propto S$
is dominant for small amplitudes $|N_2|$ in eq.(\ref{loopB3}), whereas 
$\sigma_{r\theta}$, which originates from the Lorentz force
$ j_{p} \times B_{\varphi} \propto S^2$,
 is dominant for large amplitudes.
The angular dependence of $l=4$ in Model T shown in Fig.~\ref{FigBSHx3} is 
explained by the dominant component $\sigma_{r\theta}$,
although the additional magnetic field is $l=2$.
The position of the maximum shear--strain 
is the geometrical center of the crust.
The distribution of the shear magnitude 
is more similar to that in Model P, whereas
the magnetic field in Model M is a combination of those in 
Models P and T, displayed in top panels.

The equilibrium models are critical,
because they exhibit the maximum strain $\sigma_{c}$.
Energy is released after the crustal fracture when the strain exceeds a threshold.
For example, we estimate the strain for Model M.
As displayed in the right panel of Fig.~\ref{FigBSHx3},
the magnitude of the shear stress peaks near $r/R = 0.99$.
The location corresponds to 
$\rho \approx 10^{12}{\rm g~cm}^{-3}$,
since the outer boundary $R$ corresponds to 
the neutron-drip density $\rho_{1} = 4 \times10^{11}{\rm g~cm}^{-3}$.
Assuming a fraction of $10^{-2}$ of the crust near the surface breaks
and that $10^{-2}$ of the total magnetic energy is released,
the energy is estimated as $\sim 2\times 10^{44}$ erg.
If the energy is supplied from the same fraction of elastic energy,  
the energy reduces to $4\times 10^{42}$ erg.
These energies are sufficient to induce the violent
phenomena observed in the magnetized neutron stars, although
there is uncertainty concerning whether
the elastic or magnetic energy is the actual source and what fraction of the total energy is released during explosions.
The energy estimate also depends on the magnetic
field configuration, which introduces another ambiguity.

Beyond the elastic limit, the crust may respond plastically instead of with cracking.
It is also important to evaluate to what region the magnetic field is rearranged, as
the subsequent magnetic field
evolution with  plastic flow is significantly different depending on the global or local nature of the flow
\citep[][]{2021MNRAS.506.3578G}.
Understanding the crustal dynamics beyond the elastic limit is critical for further discussions.
%

\subsection{Model comparison}

\begin{figure}
\centering
\includegraphics[scale=1.2]{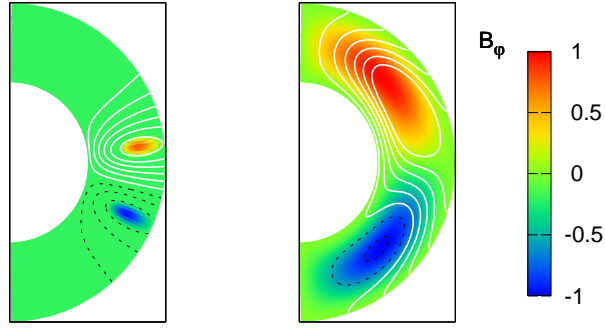}
\caption{Magnetic field for
barotropic MHD model (left) and elastic model (right) for the crust enlarged five times.
Dotted lines represent the negative value of the contour level. 
Color contour represents the toroidal component normalized by the maximum value.
}
\label{fig:mhdvx}
\end{figure}

We consider the effect of the elasticity on the equilibrium models with
mixed poloidal--toroidal magnetic fields.
In the barotoropic MHD approximation, 
 the current function $S$ is a function of
 $\Psi$ as discussed in eq.(\ref{MHDeqil.eqn}), 
and is selected as follows:
\begin{eqnarray}
   S&=& S_{0}\Psi (\Psi/\Psi_{\rm max}-1)\Theta( \Psi/\Psi_{\rm max}-1) 
    ~~{\rm for}~~ \Psi \ge 0
\nonumber
\\
    S&=& S_{0}\Psi (\Psi/\Psi_{\rm min}-1)\Theta( \Psi/\Psi_{\rm min}-1) 
    ~~~{\rm for}~~ \Psi < 0,
    \label{formSG}
\end{eqnarray}
where $\Theta$ is the usual Heaviside step function, and
$S_{0}$ is a constant.
A similar choice has been performed in previous studies,
\citep[e.g.,][]{2006ApJ...651..462Y,2009MNRAS.397..913C}
and the numerical method to calculate the equilibrium model
is discussed.
We assume that $\Psi_{\rm min}<0$ and $\Psi_{\rm max}>0$.
The toroidal magnetic field is allowed in two spatially detached regions:  $B_{\varphi}\le 0$ for $\Psi \le \Psi_{\rm min}$
and $B_{\varphi}\ge 0$ for $\Psi \ge \Psi_{\rm max}$,
while $B_{\varphi} =0$ for the intermediate. 
The toroidal field is described by $l=2$ and higher multipoles.
The magnetic function $\Psi$ for
$\Psi_{\rm min} \le \Psi \le \Psi_{\rm max}$
is smoothly connected to the exterior field, which is described by
 $l=1, 2$ and higher multipoles.

As discussed in Section 1, 
obtaining numerical models
with large toroidal magnetic-energy is difficult.
The model displayed in the left panel of Fig.~\ref{fig:mhdvx}
corresponds to the maximum of the toroidal energy.
The equilibrium model with the elastic force
is presented in the right panel of Fig.~\ref{fig:mhdvx}, which is Model M
with $l=2$, and the ratio of the field strengths is 
$(B_{\varphi})_{\rm max}/(B_{\rm p})_{\rm max}=2$.
In the barotropic model, the toroidal field is spatially localized 
inside a tiny loop in a meridian plane.
In contrast to the barotoropic MHD model,
the toroidal field extends over the crust
in the magneto--elastic equilibrium model.
Thus, a large amount of toroidal magnetic energy is stored.
%

The comparison between the two models is also summarized
via the maximum ratio of the toroidal to poloidal field strengths and energy 
in Table~\ref{table2:mylabel}.
Elasticity 
plays a crucial role in sustaining the toroidal magnetic field.

\begin{table}
    \centering
    \begin{tabular}{ccc} \hline
    Model     &  
    $(B_{\varphi})_{\rm max}/(B_{\rm p})_{\rm max}
    $ &
    $E_{\varphi}/(E_{\rm p}+E_{\varphi})$ 
\\ \hline 
   Barotropic MHD  & $9.6\times 10^{-2}$ &  $6.5\times 10^{-3}$\\
   Elastic Model &  2.0  & 0.84\\
\hline
    \end{tabular}
    \caption{Comparison of models}
    \label{table2:mylabel}
\end{table}

\section{Concluding remarks}

In this study, we investigated the equilibrium state of a magnetized neutron star crust, where the shear modulus is proportional to the density.
We focused on the magnetic field hidden in the crust, 
with a field strength that is considerably higher than that of the surface dipole field.
The magnetic field geometry containing strong poloidal and toroidal components is not obtained numerically by a barotropic MHD approximation, 
although it is conjectured to explain observational phenomena occurring in neutron stars.
Assuming that the irrotational part of the Lorentz force is balanced by the pressure and gravity,
possible equilibrium models with strong fields confined by the elastic force were demonstrated.
The solenoidal part of the Lorentz force is typically smaller than the irrotational part, and therefore the weak elastic force is sufficient for the static model.
The magnetic field configuration is less constrained, 
and a large amount of magnetic energy up to $\sim 10^{47}$ erg is sustained by the elastic force.
For example, the average field strength $B$ in the crust is
$B \sim 3\times 10^{14}$ G for $E_{\rm mag}=10^{46}$ erg, and a strong field ($>10^{14}$ G) may be hidden even when 
the surface dipole is low $<10^{13}$ G.
This case corresponds to low-field magnetars.
However, the elastic force is less effective for
a typical magnetar with a surface field strength of $\sim 10^{14}$ G.
A small deformation is allowed within a critical limit,
and the magnetic field geometry in equilibrium does not
deviate from that without the elastic force.
%

When the magnetic energy exceeds several times $ 10^{47}$ erg, the magnetic force of $B\sim 10^{15}$ G
becomes comparable to the magnitudes of the pressure and gravity at the surface.
The maximum energy stored in the crust depends considerably on the magnetic field geometry;
a larger energy is stored when the field is concentrated toward the core
because other forces also increase.
Magnetic energy in the range of 
 $2.5\times 10^{45} - 4\times 10^{48}$ erg stored in the crust
 with the low surface field of $ \sim 10^{11}$ G 
 (average field strength of $2\times 10^{14} - 5\times10^{15}$ G)
has been considered as a model for the CCO power
\citep{2020MNRAS.495.1692G,2021ApJ...909..101I}.
However, the large parameters adopted may not provide dynamically stable confinement, 
although our result for an axisymmetric model does not apply to the 3D magnetic fields considered previously.

The critical structure at the elastic limit is calculated by using a realistic shear modulus.
Correct identification of the crust-fracture location is crucial for 
the secular time evolution of the crustal magnetic field.
Elastic deformation is induced during the evolution, and 
evolution beyond a threshold is discontinuous.
Such a transition may appear as a burst in a magnetar.
The magnetic field at the critical limit is rearranged
to model the burst in numerical simulations
\citep{2011ApJ...741..123P,2020ApJ...902L..32D},
where the magnetic stress is used to estimate the critical state.
Our results show that the maximum shear--strain occurs 
near the surface
 \footnote{
The outer boundary of our model is the interface of the inner and outer crusts.  
When the toroidal magnetic field is extended to the outer crust, the region becomes 
susceptible to cracking because of the lower shear modulus.  
The total magnetic energy sustained by the elastic force is therefore reduced. 
Conversely, when the field is confined in deep interior regions,
the magnetic energy increases. 
These features can be discussed in detail only after the magnetic field configuration is clarified. 
}.
The thickness of the skin layer is
 $\sim 10^{2}$ m.
The magnetic energy stored there is
$\sim 10^{44}$ erg, and
the elastic energy is $\sim 10^{42}$ erg.
Thus, a difference of two orders of magnitude exists between two energy scales, and it is unclear which energy scale is more relevant to the observed bursts or flares.
Partial energy of this order is released during abrupt events,
but more precise investigations are required to clarify this. 

The estimates strongly depend on the field geometry.
The model considered in this study is simple, and the
magnetic field is described by a lower multipole extended to the entire crust.
Small energy scales, such as $< 10^{41}$ erg, in short bursts
\citep{2015RPPh...78k6901T} are related to
locally irregular fields, which may be concentrated elsewhere in the interior.
Irrespective of the geometry,
the effect of the elasticity considered in this study plays a crucial role in magnetic field confinement of $ \sim 10^{14}$ G in the solid crust. 

 \section*{Acknowledgements}
%
This study was supported by JSPS KAKENHI Grant Number 
JP17H06361, JP19K03850(YK), JP18H01246, JP19K14712,
JP21H01078(SK), JP20H04728(KF).
%

 \section*{DATA AVAILABILITY}
%
The data underlying this article will be shared on reasonable request 
to the corresponding author.
%

 \bibliographystyle{mnras}
 \bibliography{kojima21Spt}

\begin{thebibliography}{}
\makeatletter
\relax
\def\mn@urlcharsother{\let\do\@makeother \do\$\do\&\do\#\do\^\do\_\do\%\do\~}
\def\mn@doi{\begingroup\mn@urlcharsother \@ifnextchar [ {\mn@doi@}
  {\mn@doi@[]}}
\def\mn@doi@[#1]#2{\def\@tempa{#1}\ifx\@tempa\@empty \href
  {http://dx.doi.org/#2} {doi:#2}\else \href {http://dx.doi.org/#2} {#1}\fi
  \endgroup}
\def\mn@eprint#1#2{\mn@eprint@#1:#2::\@nil}
\def\mn@eprint@arXiv#1{\href {http://arxiv.org/abs/#1} {{\tt arXiv:#1}}}
\def\mn@eprint@dblp#1{\href {http://dblp.uni-trier.de/rec/bibtex/#1.xml}
  {dblp:#1}}
\def\mn@eprint@#1:#2:#3:#4\@nil{\def\@tempa {#1}\def\@tempb {#2}\def\@tempc
  {#3}\ifx \@tempc \@empty \let \@tempc \@tempb \let \@tempb \@tempa \fi \ifx
  \@tempb \@empty \def\@tempb {arXiv}\fi \@ifundefined
  {mn@eprint@\@tempb}{\@tempb:\@tempc}{\expandafter \expandafter \csname
  mn@eprint@\@tempb\endcsname \expandafter{\@tempc}}}

\bibitem[\protect\citeauthoryear{{Akg{\"u}n}, {Reisenegger}, {Mastrano}  \&
  {Marchant}}{{Akg{\"u}n} et~al.}{2013}]{2013MNRAS.433.2445A}
{Akg{\"u}n} T.,  {Reisenegger} A.,  {Mastrano} A.,   {Marchant} P.,  2013,
  \mn@doi [\mnras] {10.1093/mnras/stt913}, \href
  {https://ui.adsabs.harvard.edu/abs/2013MNRAS.433.2445A} {433, 2445}

\bibitem[\protect\citeauthoryear{{Archibald}, {Kaspi}, {Tendulkar}  \&
  {Scholz}}{{Archibald} et~al.}{2016}]{2016ApJ...829L..21A}
{Archibald} R.~F.,  {Kaspi} V.~M.,  {Tendulkar} S.~P.,   {Scholz} P.,  2016,
  \mn@doi [\apjl] {10.3847/2041-8205/829/1/L21}, \href
  {https://ui.adsabs.harvard.edu/abs/2016ApJ...829L..21A} {829, L21}

\bibitem[\protect\citeauthoryear{{Armaza}, {Reisenegger}  \&
  {Valdivia}}{{Armaza} et~al.}{2015}]{2015ApJ...802..121A}
{Armaza} C.,  {Reisenegger} A.,   {Valdivia} J.~A.,  2015, \mn@doi [\apj]
  {10.1088/0004-637X/802/2/121}, \href
  {https://ui.adsabs.harvard.edu/abs/2015ApJ...802..121A} {802, 121}

\bibitem[\protect\citeauthoryear{{Baiko} \& {Chugunov}}{{Baiko} \&
  {Chugunov}}{2018}]{2018MNRAS.480.5511B}
{Baiko} D.~A.,  {Chugunov} A.~I.,  2018, \mn@doi [\mnras]
  {10.1093/mnras/sty2259}, \href
  {https://ui.adsabs.harvard.edu/abs/2018MNRAS.480.5511B} {480, 5511}

\bibitem[\protect\citeauthoryear{{Bera}, {Jones}  \& {Andersson}}{{Bera}
  et~al.}{2020}]{2020MNRAS.499.2636B}
{Bera} P.,  {Jones} D.~I.,   {Andersson} N.,  2020, \mn@doi [\mnras]
  {10.1093/mnras/staa3015}, \href
  {https://ui.adsabs.harvard.edu/abs/2020MNRAS.499.2636B} {499, 2636}

\bibitem[\protect\citeauthoryear{{Braithwaite}}{{Braithwaite}}{2009}]{2009MNRAS.397..763B}
{Braithwaite} J.,  2009, \mn@doi [\mnras] {10.1111/j.1365-2966.2008.14034.x},
  \href {https://ui.adsabs.harvard.edu/abs/2009MNRAS.397..763B} {397, 763}

\bibitem[\protect\citeauthoryear{{Braithwaite} \& {Nordlund}}{{Braithwaite} \&
  {Nordlund}}{2006}]{2006A&A...450.1077B}
{Braithwaite} J.,  {Nordlund} {\r{A}}.,  2006, \mn@doi [\aap]
  {10.1051/0004-6361:20041980}, \href
  {https://ui.adsabs.harvard.edu/abs/2006A&A...450.1077B} {450, 1077}

\bibitem[\protect\citeauthoryear{{Braithwaite} \& {Spruit}}{{Braithwaite} \&
  {Spruit}}{2004}]{2004Natur.431..819B}
{Braithwaite} J.,  {Spruit} H.~C.,  2004, \mn@doi [\nat] {10.1038/nature02934},
  \href {https://ui.adsabs.harvard.edu/abs/2004Natur.431..819B} {431, 819}

\bibitem[\protect\citeauthoryear{{Caplan}, {Schneider}  \& {Horowitz}}{{Caplan}
  et~al.}{2018}]{2018PhRvL.121m2701C}
{Caplan} M.~E.,  {Schneider} A.~S.,   {Horowitz} C.~J.,  2018, \mn@doi [\prl]
  {10.1103/PhysRevLett.121.132701}, \href
  {https://ui.adsabs.harvard.edu/abs/2018PhRvL.121m2701C} {121, 132701}

\bibitem[\protect\citeauthoryear{{Chamel} \& {Haensel}}{{Chamel} \&
  {Haensel}}{2008}]{2008LRR....11...10C}
{Chamel} N.,  {Haensel} P.,  2008, \mn@doi [Living Reviews in Relativity]
  {10.12942/lrr-2008-10}, \href
  {https://ui.adsabs.harvard.edu/abs/2008LRR....11...10C} {11, 10}

\bibitem[\protect\citeauthoryear{{Ciolfi}, {Ferrari}, {Gualtieri}  \&
  {Pons}}{{Ciolfi} et~al.}{2009}]{2009MNRAS.397..913C}
{Ciolfi} R.,  {Ferrari} V.,  {Gualtieri} L.,   {Pons} J.~A.,  2009, \mn@doi
  [\mnras] {10.1111/j.1365-2966.2009.14990.x}, \href
  {https://ui.adsabs.harvard.edu/abs/2009MNRAS.397..913C} {397, 913}

\bibitem[\protect\citeauthoryear{{Colaiuda}, {Ferrari}, {Gualtieri}  \&
  {Pons}}{{Colaiuda} et~al.}{2008}]{2008MNRAS.385.2080C}
{Colaiuda} A.,  {Ferrari} V.,  {Gualtieri} L.,   {Pons} J.~A.,  2008, \mn@doi
  [\mnras] {10.1111/j.1365-2966.2008.12966.x}, \href
  {https://ui.adsabs.harvard.edu/abs/2008MNRAS.385.2080C} {385, 2080}

\bibitem[\protect\citeauthoryear{{De Grandis}, {Taverna}, {Turolla}, {Gnarini},
  {Popov}, {Zane}  \& {Wood}}{{De Grandis} et~al.}{2021}]{2021ApJ...914..118D}
{De Grandis} D.,  {Taverna} R.,  {Turolla} R.,  {Gnarini} A.,  {Popov} S.~B.,
  {Zane} S.,   {Wood} T.~S.,  2021, \mn@doi [\apj] {10.3847/1538-4357/abfdac},
  \href {https://ui.adsabs.harvard.edu/abs/2021ApJ...914..118D} {914, 118}

\bibitem[\protect\citeauthoryear{{De Luca}}{{De
  Luca}}{2017}]{2017JPhCS.932a2006D}
{De Luca} A.,  2017, \mn@doi [J. Phys.] {10.1088/1742-6596/932/1/012006}, \href
  {https://ui.adsabs.harvard.edu/abs/2017JPhCS.932a2006D} {932, 012006}

\bibitem[\protect\citeauthoryear{{Dehman}, {Vigan{\`o}}, {Rea}, {Pons}, {Perna}
   \& {Garcia-Garcia}}{{Dehman} et~al.}{2020}]{2020ApJ...902L..32D}
{Dehman} C.,  {Vigan{\`o}} D.,  {Rea} N.,  {Pons} J.~A.,  {Perna} R.,
  {Garcia-Garcia} A.,  2020, \mn@doi [\apjl] {10.3847/2041-8213/abbda9}, \href
  {https://ui.adsabs.harvard.edu/abs/2020ApJ...902L..32D} {902, L32}

\bibitem[\protect\citeauthoryear{{Duez} \& {Mathis}}{{Duez} \&
  {Mathis}}{2010}]{2010A&A...517A..58D}
{Duez} V.,  {Mathis} S.,  2010, \mn@doi [\aap] {10.1051/0004-6361/200913496},
  \href {https://ui.adsabs.harvard.edu/abs/2010A&A...517A..58D} {517, A58}

\bibitem[\protect\citeauthoryear{{Duez}, {Braithwaite}  \& {Mathis}}{{Duez}
  et~al.}{2010}]{2010ApJ...724L..34D}
{Duez} V.,  {Braithwaite} J.,   {Mathis} S.,  2010, \mn@doi [\apjl]
  {10.1088/2041-8205/724/1/L34}, \href
  {https://ui.adsabs.harvard.edu/abs/2010ApJ...724L..34D} {724, L34}

\bibitem[\protect\citeauthoryear{Enoto, Kisaka  \& Shibata}{Enoto
  et~al.}{2019}]{2019RPPh...82j6901E}
Enoto T.,  Kisaka S.,   Shibata S.,  2019, \mn@doi [Reports on Progress in
  Physics] {10.1088/1361-6633/ab3def}, 82, 106901

\bibitem[\protect\citeauthoryear{{Fujisawa} \& {Eriguchi}}{{Fujisawa} \&
  {Eriguchi}}{2013}]{2013MNRAS.432.1245F}
{Fujisawa} K.,  {Eriguchi} Y.,  2013, \mn@doi [\mnras] {10.1093/mnras/stt541},
  \href {https://ui.adsabs.harvard.edu/abs/2013MNRAS.432.1245F} {432, 1245}

\bibitem[\protect\citeauthoryear{{Fujisawa} \& {Kisaka}}{{Fujisawa} \&
  {Kisaka}}{2014}]{2014MNRAS.445.2777F}
{Fujisawa} K.,  {Kisaka} S.,  2014, \mn@doi [\mnras] {10.1093/mnras/stu1911},
  \href {https://ui.adsabs.harvard.edu/abs/2014MNRAS.445.2777F} {445, 2777}

\bibitem[\protect\citeauthoryear{{Glampedakis}, {Andersson}  \&
  {Lander}}{{Glampedakis} et~al.}{2012}]{2012MNRAS.420.1263G}
{Glampedakis} K.,  {Andersson} N.,   {Lander} S.~K.,  2012, \mn@doi [\mnras]
  {10.1111/j.1365-2966.2011.20112.x}, \href
  {https://ui.adsabs.harvard.edu/abs/2012MNRAS.420.1263G} {420, 1263}

\bibitem[\protect\citeauthoryear{{Gotthelf}, {Halpern}  \& {Alford}}{{Gotthelf}
  et~al.}{2013}]{2013ApJ...765...58G}
{Gotthelf} E.~V.,  {Halpern} J.~P.,   {Alford} J.,  2013, \mn@doi [\apj]
  {10.1088/0004-637X/765/1/58}, \href
  {https://ui.adsabs.harvard.edu/abs/2013ApJ...765...58G} {765, 58}

\bibitem[\protect\citeauthoryear{{Gourgouliatos} \& {Lander}}{{Gourgouliatos}
  \& {Lander}}{2021}]{2021MNRAS.506.3578G}
{Gourgouliatos} K.~N.,  {Lander} S.~K.,  2021, \mn@doi [\mnras]
  {10.1093/mnras/stab1869}, \href
  {https://ui.adsabs.harvard.edu/abs/2021MNRAS.506.3578G} {506, 3578}

\bibitem[\protect\citeauthoryear{{Gourgouliatos}, {Cumming}, {Reisenegger},
  {Armaza}, {Lyutikov}  \& {Valdivia}}{{Gourgouliatos}
  et~al.}{2013}]{2013MNRAS.434.2480G}
{Gourgouliatos} K.~N.,  {Cumming} A.,  {Reisenegger} A.,  {Armaza} C.,
  {Lyutikov} M.,   {Valdivia} J.~A.,  2013, \mn@doi [\mnras]
  {10.1093/mnras/stt1195}, \href
  {https://ui.adsabs.harvard.edu/abs/2013MNRAS.434.2480G} {434, 2480}

\bibitem[\protect\citeauthoryear{{Gourgouliatos}, {Hollerbach}  \&
  {Igoshev}}{{Gourgouliatos} et~al.}{2020}]{2020MNRAS.495.1692G}
{Gourgouliatos} K.~N.,  {Hollerbach} R.,   {Igoshev} A.~P.,  2020, \mn@doi
  [\mnras] {10.1093/mnras/staa1295}, \href
  {https://ui.adsabs.harvard.edu/abs/2020MNRAS.495.1692G} {495, 1692}

\bibitem[\protect\citeauthoryear{{Horowitz} \& {Kadau}}{{Horowitz} \&
  {Kadau}}{2009}]{2009PhRvL.102s1102H}
{Horowitz} C.~J.,  {Kadau} K.,  2009, \mn@doi [\prl]
  {10.1103/PhysRevLett.102.191102}, \href
  {https://ui.adsabs.harvard.edu/abs/2009PhRvL.102s1102H} {102, 191102}

\bibitem[\protect\citeauthoryear{{Igoshev}, {Gourgouliatos}, {Hollerbach}  \&
  {Wood}}{{Igoshev} et~al.}{2021}]{2021ApJ...909..101I}
{Igoshev} A.~P.,  {Gourgouliatos} K.~N.,  {Hollerbach} R.,   {Wood} T.~S.,
  2021, \mn@doi [\apj] {10.3847/1538-4357/abde3e}, \href
  {https://ui.adsabs.harvard.edu/abs/2021ApJ...909..101I} {909, 101}

\bibitem[\protect\citeauthoryear{{Kaspi} \& {Beloborodov}}{{Kaspi} \&
  {Beloborodov}}{2017}]{2017ARA&A..55..261K}
{Kaspi} V.~M.,  {Beloborodov} A.~M.,  2017, \mn@doi [\araa]
  {10.1146/annurev-astro-081915-023329}, \href
  {https://ui.adsabs.harvard.edu/abs/2017ARA&A..55..261K} {55, 261}

\bibitem[\protect\citeauthoryear{{Kojima} \& {Suzuki}}{{Kojima} \&
  {Suzuki}}{2020}]{2020MNRAS.494.3790K}
{Kojima} Y.,  {Suzuki} K.,  2020, \mn@doi [\mnras] {10.1093/mnras/staa1045},
  \href {https://ui.adsabs.harvard.edu/abs/2020MNRAS.494.3790K} {494, 3790}

\bibitem[\protect\citeauthoryear{{Kojima}, {Kisaka}  \& {Fujisawa}}{{Kojima}
  et~al.}{2021}]{2021MNRAS.506.3936K}
{Kojima} Y.,  {Kisaka} S.,   {Fujisawa} K.,  2021, \mn@doi [\mnras]
  {10.1093/mnras/stab1848}, \href
  {https://ui.adsabs.harvard.edu/abs/2021MNRAS.506.3936K} {506, 3936}

\bibitem[\protect\citeauthoryear{{Lander} \& {Gourgouliatos}}{{Lander} \&
  {Gourgouliatos}}{2019}]{2019MNRAS.486.4130L}
{Lander} S.~K.,  {Gourgouliatos} K.~N.,  2019, \mn@doi [\mnras]
  {10.1093/mnras/stz1042}, \href
  {https://ui.adsabs.harvard.edu/abs/2019MNRAS.486.4130L} {486, 4130}

\bibitem[\protect\citeauthoryear{{Lander} \& {Jones}}{{Lander} \&
  {Jones}}{2009}]{2009MNRAS.395.2162L}
{Lander} S.~K.,  {Jones} D.~I.,  2009, \mn@doi [\mnras]
  {10.1111/j.1365-2966.2009.14667.x}, \href
  {https://ui.adsabs.harvard.edu/abs/2009MNRAS.395.2162L} {395, 2162}

\bibitem[\protect\citeauthoryear{{Lander} \& {Jones}}{{Lander} \&
  {Jones}}{2011a}]{2011MNRAS.412.1394L}
{Lander} S.~K.,  {Jones} D.~I.,  2011a, \mn@doi [\mnras]
  {10.1111/j.1365-2966.2010.17998.x}, \href
  {https://ui.adsabs.harvard.edu/abs/2011MNRAS.412.1394L} {412, 1394}

\bibitem[\protect\citeauthoryear{{Lander} \& {Jones}}{{Lander} \&
  {Jones}}{2011b}]{2011MNRAS.412.1730L}
{Lander} S.~K.,  {Jones} D.~I.,  2011b, \mn@doi [\mnras]
  {10.1111/j.1365-2966.2010.18009.x}, \href
  {https://ui.adsabs.harvard.edu/abs/2011MNRAS.412.1730L} {412, 1730}

\bibitem[\protect\citeauthoryear{{Lander} \& {Jones}}{{Lander} \&
  {Jones}}{2012}]{2012MNRAS.424..482L}
{Lander} S.~K.,  {Jones} D.~I.,  2012, \mn@doi [\mnras]
  {10.1111/j.1365-2966.2012.21213.x}, \href
  {https://ui.adsabs.harvard.edu/abs/2012MNRAS.424..482L} {424, 482}

\bibitem[\protect\citeauthoryear{{Malvern}}{{Malvern}}{1969}]{1969imcm.book.....M}
{Malvern} L.~E.,  1969, {Introduction to the mechanics of a continuous medium}.
{Englewood Cliffs, N.J. : Prentice-Hall}

\bibitem[\protect\citeauthoryear{{Markey} \& {Tayler}}{{Markey} \&
  {Tayler}}{1973}]{1973MNRAS.163...77M}
{Markey} P.,  {Tayler} R.~J.,  1973, \mn@doi [\mnras] {10.1093/mnras/163.1.77},
  \href {https://ui.adsabs.harvard.edu/abs/1973MNRAS.163...77M} {163, 77}

\bibitem[\protect\citeauthoryear{{McLaughlin} et~al.,}{{McLaughlin}
  et~al.}{2003}]{2003ApJ...591L.135M}
{McLaughlin} M.~A.,  et~al., 2003, \mn@doi [\apjl] {10.1086/377212}, \href
  {https://ui.adsabs.harvard.edu/abs/2003ApJ...591L.135M} {591, L135}

\bibitem[\protect\citeauthoryear{{Mitchell}, {Braithwaite}, {Reisenegger},
  {Spruit}, {Valdivia}  \& {Langer}}{{Mitchell}
  et~al.}{2015}]{2015MNRAS.447.1213M}
{Mitchell} J.~P.,  {Braithwaite} J.,  {Reisenegger} A.,  {Spruit} H.,
  {Valdivia} J.~A.,   {Langer} N.,  2015, \mn@doi [\mnras]
  {10.1093/mnras/stu2514}, \href
  {https://ui.adsabs.harvard.edu/abs/2015MNRAS.447.1213M} {447, 1213}

\bibitem[\protect\citeauthoryear{{Pons} \& {Perna}}{{Pons} \&
  {Perna}}{2011}]{2011ApJ...741..123P}
{Pons} J.~A.,  {Perna} R.,  2011, \mn@doi [\apj] {10.1088/0004-637X/741/2/123},
  \href {https://ui.adsabs.harvard.edu/abs/2011ApJ...741..123P} {741, 123}

\bibitem[\protect\citeauthoryear{{Rea} et~al.,}{{Rea}
  et~al.}{2012}]{2012ApJ...754...27R}
{Rea} N.,  et~al., 2012, \mn@doi [\apj] {10.1088/0004-637X/754/1/27}, \href
  {https://ui.adsabs.harvard.edu/abs/2012ApJ...754...27R} {754, 27}

\bibitem[\protect\citeauthoryear{{Rea} et~al.,}{{Rea}
  et~al.}{2013}]{2013ApJ...770...65R}
{Rea} N.,  et~al., 2013, \mn@doi [\apj] {10.1088/0004-637X/770/1/65}, \href
  {https://ui.adsabs.harvard.edu/abs/2013ApJ...770...65R} {770, 65}

\bibitem[\protect\citeauthoryear{{Rea}, {Vigan{\`o}}, {Israel}, {Pons}  \&
  {Torres}}{{Rea} et~al.}{2014}]{2014ApJ...781L..17R}
{Rea} N.,  {Vigan{\`o}} D.,  {Israel} G.~L.,  {Pons} J.~A.,   {Torres} D.~F.,
  2014, \mn@doi [\apjl] {10.1088/2041-8205/781/1/L17}, \href
  {https://ui.adsabs.harvard.edu/abs/2014ApJ...781L..17R} {781, L17}

\bibitem[\protect\citeauthoryear{{Reisenegger}}{{Reisenegger}}{2009}]{2009A&A...499..557R}
{Reisenegger} A.,  2009, \mn@doi [\aap] {10.1051/0004-6361/200810895}, \href
  {https://ui.adsabs.harvard.edu/abs/2009A&A...499..557R} {499, 557}

\bibitem[\protect\citeauthoryear{{Sur}, {Haskell}  \& {Kuhn}}{{Sur}
  et~al.}{2020}]{2020MNRAS.495.1360S}
{Sur} A.,  {Haskell} B.,   {Kuhn} E.,  2020, \mn@doi [\mnras]
  {10.1093/mnras/staa1212}, \href
  {https://ui.adsabs.harvard.edu/abs/2020MNRAS.495.1360S} {495, 1360}

\bibitem[\protect\citeauthoryear{{Tayler}}{{Tayler}}{1973}]{1973MNRAS.161..365T}
{Tayler} R.~J.,  1973, \mn@doi [\mnras] {10.1093/mnras/161.4.365}, \href
  {https://ui.adsabs.harvard.edu/abs/1973MNRAS.161..365T} {161, 365}

\bibitem[\protect\citeauthoryear{{Tayler}}{{Tayler}}{1980}]{1980MNRAS.191..151T}
{Tayler} R.~J.,  1980, \mn@doi [\mnras] {10.1093/mnras/191.1.151}, \href
  {https://ui.adsabs.harvard.edu/abs/1980MNRAS.191..151T} {191, 151}

\bibitem[\protect\citeauthoryear{{Tomimura} \& {Eriguchi}}{{Tomimura} \&
  {Eriguchi}}{2005}]{2005MNRAS.359.1117T}
{Tomimura} Y.,  {Eriguchi} Y.,  2005, \mn@doi [\mnras]
  {10.1111/j.1365-2966.2005.08967.x}, \href
  {https://ui.adsabs.harvard.edu/abs/2005MNRAS.359.1117T} {359, 1117}

\bibitem[\protect\citeauthoryear{{Turolla}, {Zane}  \& {Watts}}{{Turolla}
  et~al.}{2015}]{2015RPPh...78k6901T}
{Turolla} R.,  {Zane} S.,   {Watts} A.~L.,  2015, \mn@doi [Reports on Progress
  in Physics] {10.1088/0034-4885/78/11/116901}, \href
  {http://ads.nao.ac.jp/abs/2015RPPh...78k6901T} {78, 116901}

\bibitem[\protect\citeauthoryear{{Ury{\={u}}}, {Yoshida}, {Gourgoulhon},
  {Markakis}, {Fujisawa}, {Tsokaros}, {Taniguchi}  \& {Eriguchi}}{{Ury{\={u}}}
  et~al.}{2019}]{2019PhRvD.100l3019U}
{Ury{\={u}}} K.,  {Yoshida} S.,  {Gourgoulhon} E.,  {Markakis} C.,  {Fujisawa}
  K.,  {Tsokaros} A.,  {Taniguchi} K.,   {Eriguchi} Y.,  2019, \mn@doi [\prd]
  {10.1103/PhysRevD.100.123019}, \href
  {https://ui.adsabs.harvard.edu/abs/2019PhRvD.100l3019U} {100, 123019}

\bibitem[\protect\citeauthoryear{{Ushomirsky}, {Cutler}  \&
  {Bildsten}}{{Ushomirsky} et~al.}{2000}]{2000MNRAS.319..902U}
{Ushomirsky} G.,  {Cutler} C.,   {Bildsten} L.,  2000, \mn@doi [\mnras]
  {10.1046/j.1365-8711.2000.03938.x}, \href
  {https://ui.adsabs.harvard.edu/abs/2000MNRAS.319..902U} {319, 902}

\bibitem[\protect\citeauthoryear{{Wright}}{{Wright}}{1973}]{1973MNRAS.162..339W}
{Wright} G.~A.~E.,  1973, \mn@doi [\mnras] {10.1093/mnras/162.4.339}, \href
  {https://ui.adsabs.harvard.edu/abs/1973MNRAS.162..339W} {162, 339}

\bibitem[\protect\citeauthoryear{{Yoshida}}{{Yoshida}}{2019}]{2019PhRvD..99h4034Y}
{Yoshida} S.,  2019, \mn@doi [\prd] {10.1103/PhysRevD.99.084034}, \href
  {https://ui.adsabs.harvard.edu/abs/2019PhRvD..99h4034Y} {99, 084034}

\bibitem[\protect\citeauthoryear{{Yoshida} \& {Eriguchi}}{{Yoshida} \&
  {Eriguchi}}{2006}]{2006ApJS..164..156Y}
{Yoshida} S.,  {Eriguchi} Y.,  2006, \mn@doi [\apjs] {10.1086/501050}, \href
  {https://ui.adsabs.harvard.edu/abs/2006ApJS..164..156Y} {164, 156}

\bibitem[\protect\citeauthoryear{{Yoshida}, {Yoshida}  \& {Eriguchi}}{{Yoshida}
  et~al.}{2006}]{2006ApJ...651..462Y}
{Yoshida} S.,  {Yoshida} S.,   {Eriguchi} Y.,  2006, \mn@doi [\apj]
  {10.1086/507513}, \href
  {https://ui.adsabs.harvard.edu/abs/2006ApJ...651..462Y} {651, 462}

\makeatother
\end{thebibliography}
\end{document}